# Cluster glass transition and relaxation in random spinel $CoGa_2O_4$

T. Naka[1,*], T. Nakane[1], S. Ishii[2], M. Nakayama[1], A. Ohmura[3], F. Ishikawa[3], A. de Visser[4] H. Abe[5], and T. Uchikoshi[1]

[1]*National Institute for Materials Science, 1-2-1, Sengen, Tsukuba, Ibaraki 305-0047, Japan*
[2]*Department of Physics, Tokyo Denki University, Hatoyama, Saitama 350-0394, Japan*
[3]*Faculty of Science, Niigata University, Niigata, 950-2181, Japan*
[4]*Van der Waals – Zeeman Institute, Institute of Physics, University of Amsterdam, Science Park 904, 1098 XH, The Netherlands*
[5]*Joint and Welding Research Institution, Osaka University, 11-1 Mihogaoka, Ibaraki, Osaka 567-0047, Japan*

**Abstract**

We report magnetic properties in the random spinel magnet $CoGa_2O_4$. Rietveld analysis of the x-ray diffraction profile for $CoGa_2O_4$ reveals that the Co and Ga ions are distributed randomly in the tetrahedral A-sites and octahedral B-sites in the cubic spinel structure. $CoGa_2O_4$ exhibits a spin-glass transition at $T_{SG} = 8.2$ K that is confirmed by measurements of the dc- and ac-susceptibilities and thermoremanent magnetization (TRM) that develops below $T_{SG}$. From the frequency dependence of the freezing temperature $T_f$ for $CoGa_2O_4$, it is indicated that the relaxation time $\tau(T)$ follows a Vogel–Fulcher law $\tau = \tau_0\exp[-E_a/k_B(T - T_0)]$. An analysis of specific heat suggested that a doublet ground state of the octahedrally coordinated $Co^{2+}$ was stabilized by spin-orbit and crystal field couplings. The relaxation rate of TRM is considerably enhanced at $T_{SG}$ and decays rapidly above and below $T_{SG}$. The time course of TRM is reproduced by non-exponential relaxation forms, such as a stretched exponential (Kohlrausch) as well as Ogielski and Weron relaxation forms. This behavior is displayed universally in glass systems, and the characteristic parameters associated with these functions were reasonable.

Key words: thermoremanent magnetization, specific heat, complex system, nonextensive parameter

*Corresponding author: naka.takashi@nims.go.jp

# I. INTRODUCTION

The A-site spinel antiferromagnets (AFMs) with stoichiometry $AB_2O_4$, where A is a divalent magnetic cation and B is a trivalent nonmagnetic cation, have been a platform for the realization of novel magnetic states (e.g., quantum spin or orbital liquid states [1–3]), pressure-induced valence transitions [4–5], and a Néel to spin-spiral (NSS) transition [2], which is characterized by a propagation vector $q$ that forms continuous surfaces in $k$-space. In the spinel structure, the A and B cations occupy tetrahedral and octahedral sites, respectively. The former (latter) has previously been called the A-site (B-site). Actually, the A-site forms a diamond lattice, which consists of two interpenetrated face-centered cubic (fcc) lattices shifted along the [111] axis. The A-site spinel is therefore magnetically bipartite. In other words, the Néel state is realized if one introduces only a nearest neighbor antiferromagnetic interaction $J_1$ between spins occupying sites in the respective fcc sublattices. As indicated in Fig. 1(a), a next-nearest-neighbor antiferromagnetic interaction $J_2$ acts competitively with the nearest neighbor interaction, and a magnetic frustration effect is therefore anticipated. For such A-site frustrated antiferromagnets, a Monte-Carlo simulation shows a novel phase diagram in the temperature-($J_2/J_1$) plane. The simulation predicts that the Néel state is destabilized rapidly with increasing $J_2/J_1$ and that a magnetic transition to a spin spiral state occurs at $J_2/J_1 = 1/8$; meanwhile, densely quasi-degenerated magnetic states characterized by propagation vector $q$ form continuous surfaces in $k$-space for $J_2/J_1 > 1/8$ (Fig. 1(b)) [2]. In the typical exemplary system, $CoAl_2O_4$ is expected to be a platform that exhibits novel magnetic behavior. Because "frustration" is an inherent topological feature of the A-site spinel or diamond lattice AFMs, the magnetic states are quite sensitive to crystallographic perturbations such as inversion [6, 7], magnetic dilution [8], and chemical and physical pressures [8, 9]. The inversion, which is a chemical disorder (antisite defect of cationic configuration between the A-sites and B-sites), is represented by the chemical formula $(A_{1-\eta}B_\eta)[B_{2-\eta}A_\eta]O_4$. In this representation, parentheses and square brackets indicate the A-site and B-site occupations, respectively, and η is the so-called inversion parameter. Microscopically, an inverted $Co^{2+}$ ion at the B-site couples strongly with neighboring $Co^{2+}$ ions at the A-site and B-site via exchange interactions $J_{AB}$ and $J_{BB}$, respectively (Fig. 1(a)).One can deduce therefore that these exchange interactions are spatially and randomly distributed in the A-site spinel and brings an exchange disorder, which seems to act as strong magnetic perturbations to the magnetic state, especially, lifting the degeneracy due to the frustration and stimulating an ordering transition [10]. It has been revealed experimentally that a Néel-to-spin-glass (NSG) transition occurs at $\eta_c \sim 0.08$ in a partially inverted $CoAl_2O_4$ [6], whereas in the magnetically diluted system $Co_{1-x}Zn_xAl_2O_4$, the transition occurs at $x_c = 0.06$ [8]. The higher sensitivity of the magnetic ground state with respect to these chemical disorders is due to the magnetic state of $CoAl_2O_4$, which is expected to be located in the vicinity of $J_2/J_1 = 1/8$ where the NSS transition occurs [2], as shown schematically in Fig, 1(b). At the phase boundary, the facts suggest that the transition temperature is drastically reduced, and concomitantly, the magnetic stiffness constant $\kappa$ becomes zero, the magnetically ordered states are destabilized at $J_2/J_1 = 1/8$. In other words, a "classical" spin liquid is realized at low temperatures in the vicinity of $J_2/J_1 = 1/8$. The exchange stiffness constant $\kappa$ is related specifically to the energy cost of magnetically ordered spins when the alignment of neighboring spins coupled with an exchange interaction is distorted infinitesimally from that of the ground state. Experimentally, the magnetic ground state has been proposed to be a spin-glass [6, 11], spin-liquid [6, 12], unconventionally magnetically ordered [13, 14], and antiferromagnetic [15, 16] state.

Recent neutron diffraction measurements by MacDougall *et al*. [17] have revealed that (i) an intense Lorentzian scattering component observed in single crystals with $\eta = 0.02 \pm 0.04$ remains even well below the freezing temperature $T^* = 6.5 \pm 1$ K, (ii) an anisotropic Lorentzian-squared character is also established below $T^*$, and (iii) the observed antiferromagnetic spin wave excitation is consistent with that of antiferromagnetism with a $J_2/J_1$ value of $0.110 \pm 0.003$, which is close to the NSS boundary $J_2/J_1 = 0.125$, but falls in the Néel ordered region (Fig. 1 (b)), where a first-order phase transition is theoretically expected to occur [2]. A similar value of $J_2/J_1 = 0.109 \pm 0.002$ has been reported by Zaharko *et al*. using a single-crystal specimen with $\eta = 0.05$ [14, 18]. Based on these observations, MacDougall *et al*. [17] have hypothesized that there is a peculiar magnetic feature below $T^*$, that is, the antiferromagnetic domain wall motion is kinetically inhibited, and the long-range antiferromagnetic correlation is therefore blocked. The incomplete magnetic order (the suppression of antiferromagnetic long-range order) found also by Zaharko *et al*. [14, 18] immediately suggests that a frozen and fragmented AFM structure is formed. Indeed, the discrepancy between magnetic susceptibilities measured after ZFC (zero field cooling) and FC (field cooling) [6, 8, 18] and the emergence of the thermoremanent magnetization (TRM) [8] observed in $CoAl_2O_4$ are indicative of small-volume (finite-size) effects and glassy natures. In addition to these phenomena, a relaxation of TRM, a smear-out of a phase transition, and exchange-bias are also indicative of nanomagnets [19] and spin-glass [20].

The magnetic states of previously examined $CoAl_2O_4$ samples have been located near both the NSS and NSG boundaries [8, 15]. This condition seems to be the underlying cause of their controversial behavior with respect to magnetism, and it may account, in particular, for the magnetic ground state and the incomplete magnetic ordering of the A-site spinel $CoAl_2O_4$. For $CoGa_2O_4$, in contrast to the case of $CoAl_2O_4$, we might expect that the magnetic states would not be near both boundaries because the degree of chemical disorder is quite sufficient to diminish the effect of magnetic frustration discussed theoretically in Ref. 2 and far from the NSG boundary of $\eta_c \sim 0.08$ [7]. Both the Co and Ga ions are distributed randomly to the A- and B-sites. Actually, the inversion parameters determined by powder neutron [21] and single-crystal x-ray [22] diffraction are 0.60 and 0.575(4), respectively. These values of $\eta$ are comparable with the value of 0.63 obtained by Molet *et al*. for a powder sample [7]. The magnetic ground state of $CoGa_2O_4$ has been postulated to be a spin-glass state below $T \sim 10$ K because the dc-magnetic susceptibility curve shows splitting between field-cooled and zero-field cooled measurements [7, 21], and ac-susceptibility shows a cusp at $T_f \sim 10$ K [21]. The freezing temperature, $T_f$, increases with increasing frequency [23]. These spin-glass type behaviors are consistent with the fact that neutron diffraction does not show any magnetic reflections at $T = 1.5$ K [21].

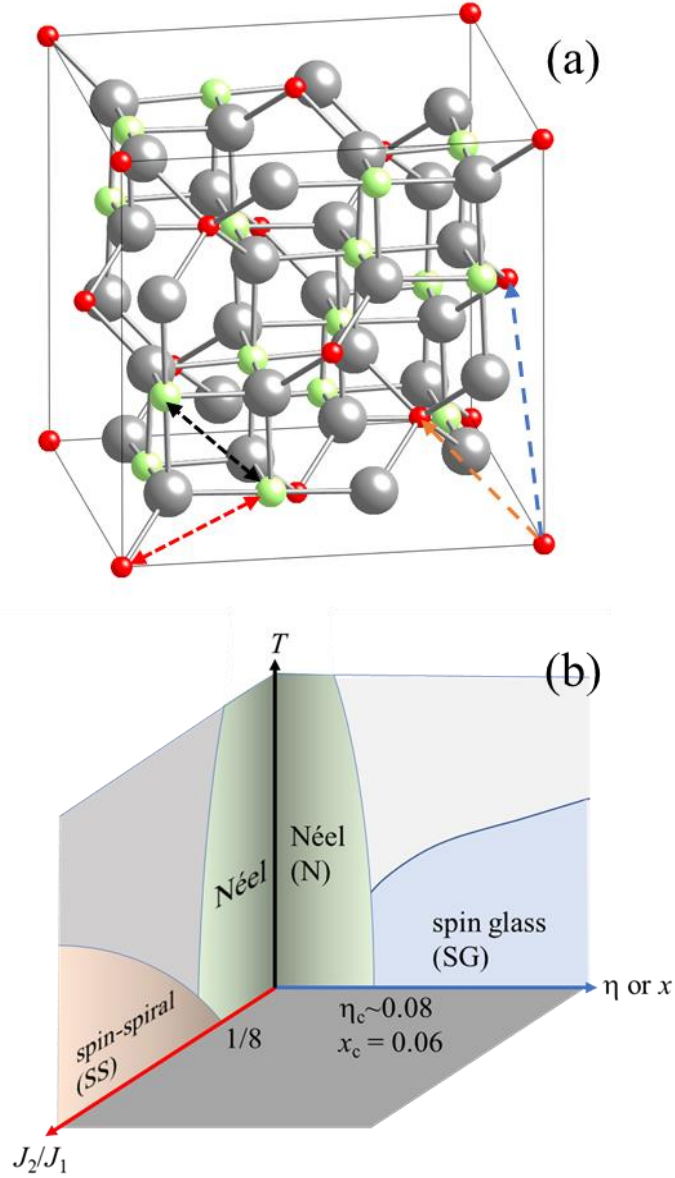

FIG. 1 (a) Spinel structure. Red, yellow, and gray spheres represent the tetrahedral, octahedral cation, and oxygen sites, respectively. Orange and blue dashed lines are connected between nearest and next-nearest-neighbor A-A pairs. Red and black dashed lines represent nearest neighbor A-B and B-B pairs, respectively. (b) Schematic phase diagrams of the $AB_2O_4$ magnet projected into the $\eta(x)$-$T$ [6, 8] and $(J_2/J_1)$-$T$ [2] planes.

In this paper, we report that comprehensive structural and magnetic investigations in the ternary spinel oxide $CoGa_2O_4$ reveal that the inversion parameter is close to that of a random spinel $\eta = 3/2$ and that the cluster spin-glass state is realized below the spin-glass transition at $T_{SG} = 8.2$ K. Remarkably, TRM is observed in $CoGa_2O_4$, and it decays non-exponentially below $T_{SG}$ as a function of time. We demonstrated successfully that a generalized relaxation function based on the stochastic theory of dielectric relaxation derived by Weron [24] was reproduced experimentally in the TRM relaxation. Interestingly, for the insulating spin-glass $CoGa_2O_4$, the parameter $q$ extracted from the TRM relaxation seemed to show a temperature dependence closely resembling that obtained by neutron spin echo measurements [25] in canonical (metallic) spin-glass systems. The parameter $q$ has been characterized as a non-extensive entropy parameter and was originally introduced in a generalization of Boltzmann–Gibbs entropy by Tsallis [26, 27]. In this study, $q$ was specifically a measure of the collective and hierarchical nature of relaxation phenomena in spin-glass systems [28]. Moreover, the relaxation time $\tau(T)$ obtained from the frequency dependence of the freezing temperature and the TRM decay curve suggested the existence of a crossover in the relaxation process at $T \sim T_{SG}$ from a high-temperature, Vogel–Fulcher type to a low-temperature, quantum mechanical relaxation process. These findings might facilitate further investigation and give insight into the characteristics of the frustrated A-site antiferromagnets, such as $CoAl_2O_4$, the magnetic ground state of which has not yet been revealed.

## II. EXPERIMENTAL

A solid–solid reaction method was used to synthesize $CoGa_2O_4$ and $ZnGa_2O_4$ polycrystalline samples with the proper amounts of CoO (4N), ZnO(4N), and $Ga_2O_3$ (4N). The mixed powder was calcinated at 1300°C in ordinary air for 24 h and cooled to room temperature at a rate of 42.5 °C /h. We performed powder x-ray diffraction measurements using a synchrotron x-ray source and made crystal structure refinements with Rietveld analysis software, RIETAN-PF [29]. Powder x-ray diffraction measurements using synchrotron radiation ($\lambda = 0.620089$ Å for $CoGa_2O_4$ and $0.65296$ Å for $ZnGa_2O_4$) were conducted on the BL15XU beam line at SPring-8 (Harima, Japan) [30]. The dc- and ac-magnetizations and specific heat at ambient pressure were measured using Magnetic Properties Measurement System (MPMS-XL; Quantum Design) and Physical Properties Measurement System (PPMS Dynacool; Quantum Design), respectively. The error in phase for ac-susceptibility was corrected by that measured for a paramagnetic reference material $Dy_2O_3$. The temperature dependence and relaxation of TRM were measured after field cooling with an excitation dc-field of $H_{FC}$ = 100 Oe applied at $T$ = 70 K. After the measurement temperature $T = T_m$ had been reached, the sample was held at that temperature for $t_m$ = 100–300 s. The magnetic field was then immediately reduced to zero. Fourier transform infrared (FT-IR) spectroscopy was carried out using a conventional spectrometer (FT/IR-6200; JASCO) with KBr pellets.

## III. RESULTS

### A. X-ray diffraction

Figures 2(a) and (b) show x-ray diffraction profiles for $CoGa_2O_4$ and $ZnGa_2O_4$, respectively. Each sample can be assigned to have a cubic spinel structure (space group: Fd-3m) with a cationic configuration disorder quantified by inversion parameter, $\eta$ = 0.664(8) for $CoGa_2O_4$. Because the difference of atomic form factors for Zn and Ga is too small to determine $\eta$ reasonably by Rietveld refinement, we assumed that the degree of inversion was zero for $ZnGa_2O_4$. The crystallographic parameters and reliable parameters refined by the Rietveld method are listed in Table I. The parameters obtained for $CoGa_2O_4$ in this work are in good agreement with previously reported values [7, 21, 22].

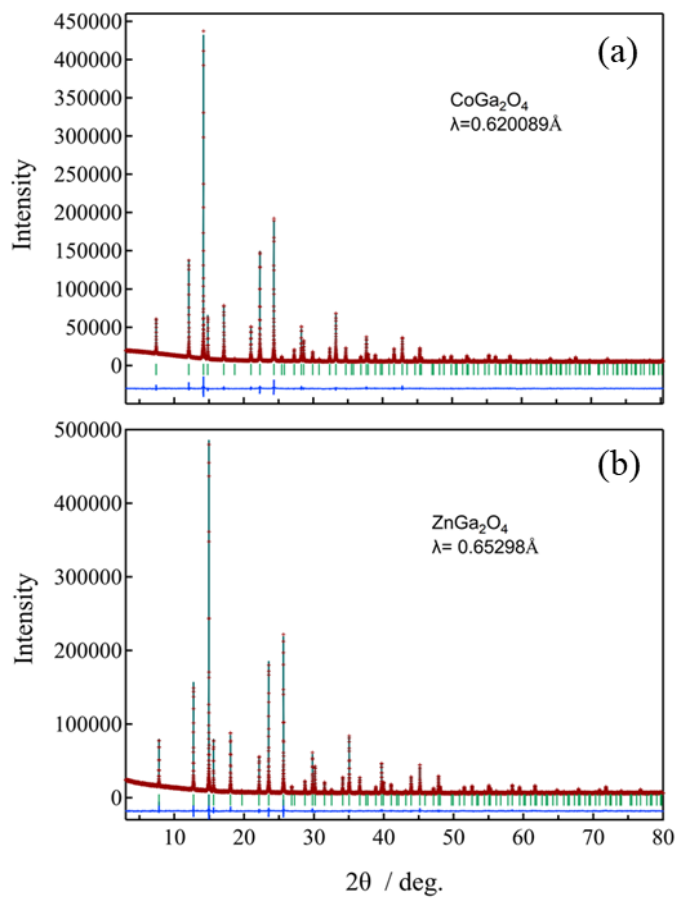

FIG. 2 X-ray diffraction profiles for (a) $CoGa_2O_4$ and (b) $ZnGa_2O_4$. Note that the x-ray wavelength used for $CoGa_2O_4$ and $ZnGa_2O_4$ is slightly different. The vertical green lines show positions of the Bragg reflections. The difference between observed and calculated intensities is plotted as the lower blue trace.

Table I Crystallographic parameters and reliability factors obtained by Rietveld refinement for CoGa$_2$O$_4$ and ZnGa$_2$O$_4$.

| Sample | $a$(Å) | $u$ | $g_A$(Co/Zn)[*] | $g_A$(Ga) | $g_B$(Co/Zn)[*] | $g_B$(Ga)[*] | $R_{wp}$ | $S$ |
|---|---|---|---|---|---|---|---|---|
| CoGa$_2$O$_4$ | 8.32654(1) | 0.25844(11) | 0.336(8) | 0.664(8) | 0.332(4) | 0.668(4) | 4.04 | 3.56 |
| CoGa$_2$O$_4$[**] | 8.325 | 0.2582 | 0.38 | 0.62 | 0.31 | 0.69 | - | - |
| ZnGa$_2$O$_4$ | 8.33361(2) | 0.26265(14) | 1.0 | 0 | 0 | 1.0 | 5.16 | 4.85 |

[*]Constraints: $g_A$(Co/Zn) + $g_A$(Ga) =1, $2g_B$(Co/Zn) =1 − $g_A$(Co/Zn), $0.5g_A$(Ga) + $g_B$(Ga) =1 for CoGa$_2$O$_4$. [**]Melot et al [7]. Assumed to be $g_A$(Zn) = 1.0 and $g_B$(Ga) = 1.0 for ZnGa$_2$O$_4$.

### B. Dc-susceptibility

Figure 3(a) shows the temperature dependence of magnetic susceptibility $\chi(T)$ after zero ZFC and FC at $H_{FC}$ = 100 Oe. The cusp exhibited by $\chi(T)$ after ZFC at $T$ = 8.5(5) K signals a spin-glass transition. As shown in Fig. 3(b), $\chi(T)$ obeys a modified Curie-Weiss law, $\chi(T) = \chi_0 + C/(T − \theta)$, where $\chi_0$, $C$, and $\theta$ are the temperature-independent susceptibility, Curie constant, and Weiss temperature, respectively. The constant susceptibility $\chi_0$ = −8.21(1) × 10$^{-5}$ emu/mol (Langevin diamagnetic susceptibility) obtained by measuring $\chi(T)$ for ZnGa$_2$O$_4$ was evaluated as the constant susceptibility for CoGa$_2$O$_4$. The parameters estimated by the least squares method are $C$ = 3.08(0) and $\theta$ = −49.9(3) K. The effective moment $p_{eff}$ = 4.97(0) $\mu_B$, which is calculated from the value of $C$, was close to reported values [7, 31] but rather larger than 4.51 $\mu_B$ for CoAl$_2$O$_4$ [8]. The observation that $p_{eff}$ was larger for CoGa$_2$O$_4$ than for CoAl$_2$O$_4$, which has a smaller inversion with $\eta$ = 0.055 [8], might have resulted from the significant occupation of the B-site ($\eta$ = 0.66) by Co$^{2+}$. The value of $p_{eff}$ was estimated with the simple equation $p_{eff}^2 = (1 − \eta)p_{eff}^2(A) + \eta p_{eff}^2(B)$, where $p_{eff}(A)$ and $p_{eff}(B)$ are the effective moments of the Co$^{2+}$ ions at the A- and B-sites, respectively. The value of $p_{eff}(B)$, 5.22 $\mu_B$, was significantly larger than that of $p_{eff}(A)$, 4.47 $\mu_B$, and the value of $g[S(S+1)]^{0.5}$ =

3.87 for $S = 3/2$ and $g = 2$. If the spin-orbit coupling is negligibly small compared with the octahedral crystal field energy, the 3d-electronic state of $Co^{2+}$ is split into doubly degenerate $d\gamma$ and triply degenerate $d\varepsilon$ states by the tetrahedral or octahedral ligand fields, but the $d\gamma$ ($d\varepsilon$) state lies below the $d\varepsilon$ ($d\gamma$) state for $Co^{2+}$ ions that occupy A-sites (B-sites). The electronic configurations of $Co^{2+}$ in the A-site and B-site can be represented with $(d\gamma)^4(d\varepsilon)^3$ and $(d\varepsilon)^5(d\gamma)^2$, respectively. The latter therefore has an orbital degree of freedom, that is, $2\upsilon + 1 = 3$, where $\upsilon = 1$ is a pseudo orbital moment for the $(d\varepsilon)^5(d\gamma)^2$ configuration. The former has no orbital degree of freedom because $\upsilon = 0$ for the $(d\gamma)^4(d\varepsilon)^3$ configuration. As mentioned below, however, the spin and orbital (configurational) degrees of freedom involved with the inverted $Co^{2+}$ ion, which was primarily associated with $CoGa_2O_4$, seemed to be reduced by spin-orbit coupling.

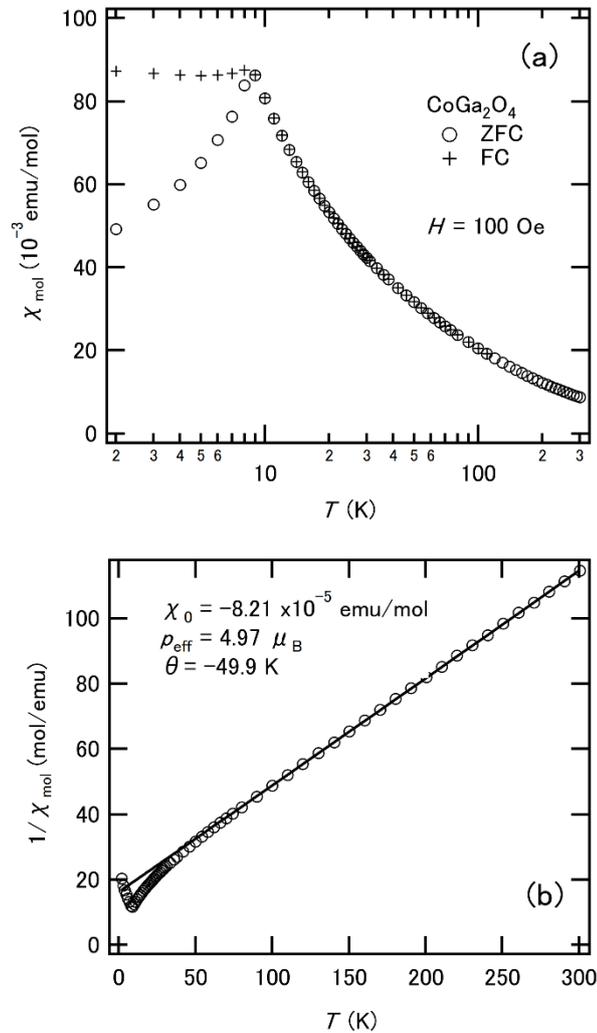

FIG. 3 Temperature dependence of (a) magnetic susceptibility $\chi_{mol}$ and (b) reciprocal susceptibility $1/\chi_{mol}$ for CoGa$_2$O$_4$.

## C. Ac-susceptibility

The spin-glass state can be confirmed by the dynamical response of the frequency dependence of ac-susceptibility within the frequency range $0.3 \leq \nu \leq 1500$ Hz in the vicinity of the transition. Figure 4 shows the real ($\chi'$) and imaginary ($\chi''$) components of the ac-susceptibility $\chi(\nu)$ as a function of temperature at various frequencies. $\chi'(T)$ exhibits a cusp at the freezing temperature $T_f$, which decreases with increasing frequency $\nu$. The maximum value of $\chi'$ (at $T = T_f$) decreases with increasing $\nu$. These frequency-dependent characteristics are typically observed in spin-glass systems. Note that in addition to the spin glass behavior around $T = 9$ K, a weak frequency dependence in $\chi'(T)$ is observable at high temperatures below 50 K. The relaxation time $\tau_{ac} = \nu^{-1}$, the mean time for spin flipping, shows a divergence due to the critical slowing down at the spin-glass transition temperature $T_{SG} = 8.2$ K (see APPENDIX A). The temperature dependence of $\tau_{ac}$ deviates remarkably from the Arrhenius law $\tau_{ac}(T) = \tau_0 \exp(E_a/k_B T)$ (Fig. 5). It is generally recognized that in a spin-glass system, magnetic clusters form that consist of strongly interacting spins blocked in random directions whose size and correlation length $\xi$ grow with decreasing temperature. Consequently, $\tau(T)$ tends to diverge at low temperatures. The deviation from the Arrhenius law suggests that a temperature-dependent activation energy can be employed. The strong temperature dependence of $\tau$ evidenced in CoGa$_2$O$_4$ can be described by an empirical equation, the so-called Vogel–Fulcher law,

$$\tau_{ac}(T) = \tau_0 \exp\left(\frac{E_a}{k_B(T_f - T_0)}\right), \tag{1}$$

where $\tau_0$, $k_B$, and $T_0$ are the characteristic time for spin-flipping, the Boltzmann constant, and the material-dependent characteristic temperature (denoted as the Vogel–Fulcher temperature), respectively (see Fig. 5). Based on the derivation of the quantities in Eq. (1) in the intermetallic spin-glass compound PrRhSn$_3$ [32], one can derive $\tau_0 = 2.86 \times 10^{-10}$ s, $E_a/k_B = 42.3(6)$ K, and $T_0 = 7.14(3)$ K for CoGa$_2$O$_4$. The value of $\tau_0$ is considerably larger than the values of typical canonical spin-glass systems (e.g., $\tau_0 \sim 10^{-12}$ s for single-spin flipping process). The slow dynamics observed in $CoGa_2O_4$ suggests that strongly interacting clusters are established in the glass state. The experimentally obtained parameters of $\Delta T_f/T_f \Delta \log \nu = 0.028$ (comparable to that of the cluster-glass system $PrRhSn_3$), $(T_f − T_0)/T_f = 0.22$ (comparable to that of a short-range–interaction spin-glass system), and $E_a/k_B T_f = 5.9$ at $\nu = 3$ Hz (comparable to those of $Eu_{0.4}Sr_{0.6}S$ and Mn-aluminosilicate) lie in the range of the empirical values for insulating, short-range–interaction spin-glass systems, and most probably cluster-glass [33]. The implication is therefore that the random spinel $CoGa_2O_4$ undergoes a cluster-glass transition at $T_{SG} = 8.2$ K.

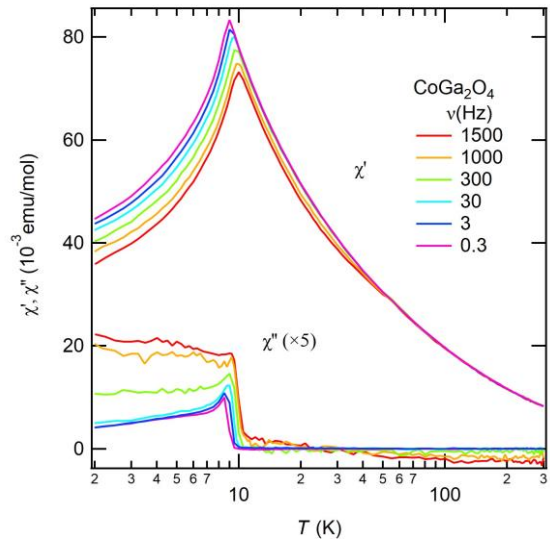

FIG. 4 Real ($\chi'$) and imaginary ($\chi''$) components of the ac-susceptibility as a function of temperature at various frequencies.

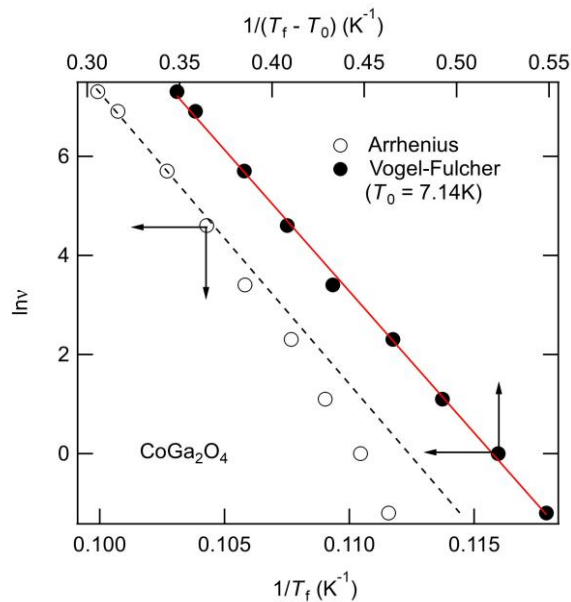

FIG. 5 Semi-logarithmic plot of frequency ν versus $1/T_f$ (Arrhenius plot) and $1/(T_f - T_0)$ (Vogel–Fulcher plot). Dashed and solid lines represent least squares fitting curves of the Arrhenius function at high frequencies and the Vogel–Fulcher function, respectively.

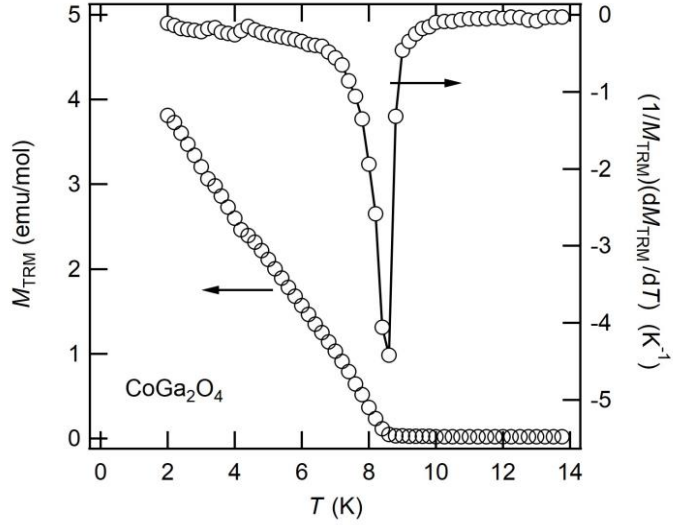

FIG. 6 $M_{TRM}$ and $(1/M_{TRM})(dM_{TRM}/dT)$ as a function of temperature.

### D. Thermoremanent magnetization

Generally, thermoremanent magnetization (TRM) is associated with spin-glasses or superparamagnetism [34-36]. Since the discovery of spin-glass TRM, it has been used as a probe of the field-quenched state in spin-glasses [35, 36] and of aging (i.e., memory and rejuvenation) effects realized in spin and also cluster glasses [37, 38]. Here we focus on both its magnetic relaxation and critical behavior in the vicinity of and below $T_{SG}$. In $CoGa_2O_4$ and also in the isostructural $CoAl_2O_4$, the spin-glass state is induced by the chemical disorder that accompanies the enhancement of TRM [8]. Actually, the value of TRM, $M_{TRM}$, increases monotonically with increasing η [39] and is enhanced anomalously at the magnetic phase boundary in a diluted system [8]. The relaxation and relaxation component ratio of TRM for $CoGa_2O_4$ can signal features of magnetic properties, as shown below. $M_{TRM}(T)$ develops rapidly below $T_{SG}$, but it is clearly apparent even slightly above $T_{SG}$ under our experimental conditions, whereas the derivative of TRM with respect to temperature shows a step-like enhancement at $T$ = 8.2 K (not shown). The logarithmic derivative, $d\ln M_{TRM}/dT$ ($\approx$ $(1/M_{TRM})(dM_{TRM}/dT)$), shows a more definitive peak at $T$ = 8.5 K, a temperature slightly higher than $T_{SG}$ (Fig. 6). Similarly, the relaxation ratio, defined as $\Delta M_{TRM}/M_{TRM} = [M_{TRM}(t_f) - M_{TRM}(t_i)]/M_{TRM}(t_i)$, which is extracted from the isothermal decay curve at various temperatures (Fig. 7(a)), exhibited a sharp peak at $T$ = 8.2 K, which corresponds to the spin-glass transition temperature $T_{SG}$ = 8.2 K obtained above (Fig. 7 (b)). Here $t_i$ and $t_f$ are the initial and final points in the decay curves, respectively. Note that one can derive $\Delta M_{TRM}/M_{TRM} = -\Delta t/\tau$ if one assumes a Debye-type relaxation $M_{TRM}(t) = M_{TRM}(0)\exp(-t/\tau_D)$, where $M_{TRM}(0)$ and $\tau_D$ are the TRM value at $t = 0$ and the relaxation time, respectively, and $\Delta t = t_f - t_i$. Below $T_{SG}$, the relaxation rate decays rapidly with decreasing temperature, and its temperature variation can be fitted by an Arrhenius function with a constant term, $\Delta M_{TRM}/M_{TRM} \sim a_0 + a_1\exp(-E_{Arr}/k_B T)$, where $a_0$ and $a_1$ are temperature-independent constants, and $E_{TRM}$ is an activation energy for the TRM relaxation (Fig. 7(b)). We extracted the quantities, $a_0 = -0.041(5)$, $a_1 = -3.5(7)$, and $E_{Arr}/k_B = 18(2)$ K by a least squares fitting to the Arrhenius function. As shown below, the value of $E_{TRM}/k_B$ was consistent with the temperature when the specific heat is a maximum. Consequently, we can roughly estimate $\tau_D(T)$ at $T$ = 2 K and $T_{SG}$ to be ~ $10^6$ and $10^4$ s, respectively. These times are one order of magnitude larger than the values obtained directly from the isothermal decay curves of $M_{TRM}$ fitted by other decay functions (see APPENDIX B). Instead, the $M_{TRM}(t)$ curve is reproduced by the so-called stretched exponential and other functional forms, as mentioned below.

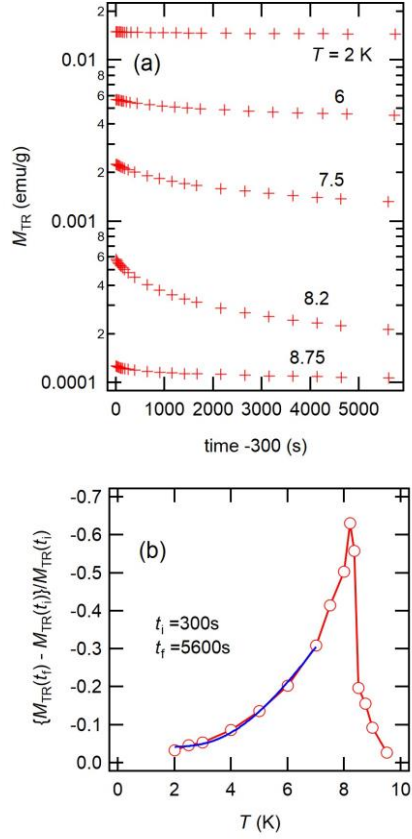

FIG. 7 (a) Decay profiles and (b) temperature dependence of the relaxation rate $[M_{TRM}(t_f) - M_{TRM}(t_i)]/M_{TRM}(t_i)$ of thermoremanent magnetization $M_{TRM}$. Red solid line is a guide to aid visualization. Blue solid line represents a fitting curve to an Arrhenius function (see text).

Figures 8(a) and (b) show isothermal curves of $M_{TRM}(t)$ at $T = 3$ ($< T_{SG}$) and 8.5 K ($> T_{SG}$), respectively, as a function of time after the excitation field ($H_{FC} = 100$ Oe) had been turned off. In advance, the sample was field-cooled with $H_{FC} = 100$ Oe from 70 K to the measurement temperature $T$. Before measuring $M_{TRM}(t)$ at the measurement temperature, the magnetic field $H_{FC} = 100$ Oe was kept for a waiting time $t_w = 300$ s. It is apparent that the profiles of $M_{TRM}(t)$ change qualitatively at $T \sim T_{SG}$ (Figs. 8(a) and (b)). Below $T_{SG}$, the variation of $M_{TRM}$ with time for $CoGa_2O_4$ follows fairly typical relaxation functions: a stretched exponential (Kohlrausch), or the Ogielski and Weron functions (see APPENDIX B) with reasonable parameters. Figure 8(a) shows the isothermal decay of $M_{TRM}$ at 3 K with least squares fits for the Weron, Ogielski, and Kohlrausch functions displayed by red, blue, and green dashed lines, respectively. The $M_{TRM}(t)$ curve is reproduced fairly well by the Weron and Ogielski functions but not as well by the Kohlrausch function. As demonstrated in APPENDIX C, the $t_w$-dependent behavior, which is one of the aging effects often apparent in TRM decay for spin and cluster glasses [36, 39], can be neglected for $CoGa_2O_4$ to facilitate extraction of the relaxation parameters for the relaxation functions from the fitted $M_{TRM}(t)$ curve. Above and in the vicinity of the spin-glass transition, a time-independent component term must be added to these functions to reproduce the $M_{TRM}(t)$ curve:

$$M_{TRM} = p_0 + p_1 \phi(t), \qquad (2)$$

where $p_0$ and $p_1$ are time-independent constants (magnetization components), and $\phi(t)$ is the relaxation function mentioned above (Fig. 8(b)). Note that the residual field $H_{res}$ in the magnetometer, which was estimated to be a few tenths Oe, can make an extrinsic contribution to the value of $p_0$, which is estimated to be $M_{res} = \chi(T \sim T_{SG})H_{res} \sim 1\times 10^{-4}$ emu/g ($\sim 3\times 10^{-2}$ emu/mol). This value is comparable to the value of $p_0$ observed at 9 K.

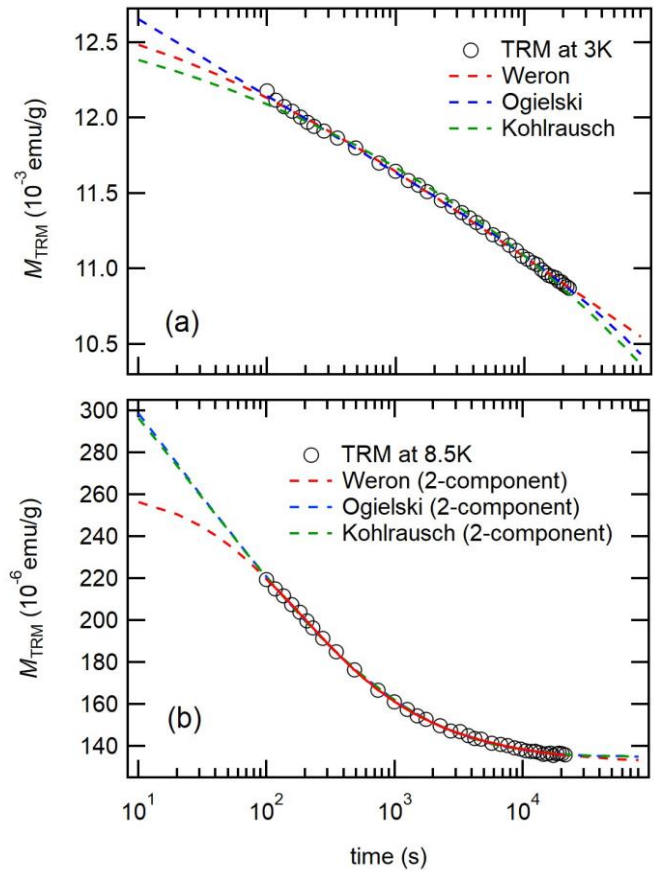

FIG. 8 Isothermal decays of $M_{TRM}$ measured at (a) $T = 2$ K and (b) 8.5 K. The red, blue, and green dashed lines represent curves obtained by least squares fitting of the Weron, Ogielski, and Kohlrausch relaxation functions, respectively.

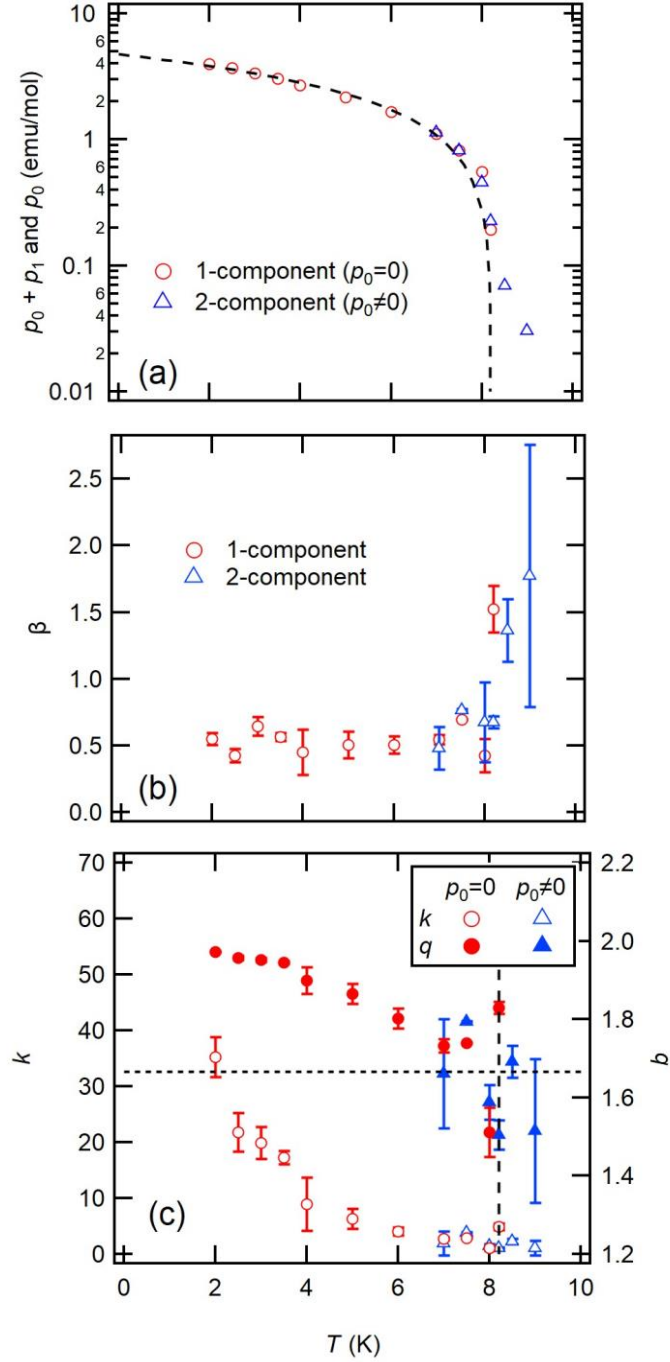

FIG. 9 Temperature dependences of (a) $p_{tot} = p_0 + p_1$, (b) $\beta$, and (c) $k$ and $q$. Dashed curve in (a) represents a numerically fitted curve that corresponds to $p_{tot}(T) = p_{tot}^0 (1 - T/T_{SG})^\alpha$ with $p_{tot}^0 = 4.4$ emu/mol and $\alpha = 0.77$. Vertical and horizontal dashed lines in (c) indicate $q = 5/3$ and $T = T_{SG}$, respectively.

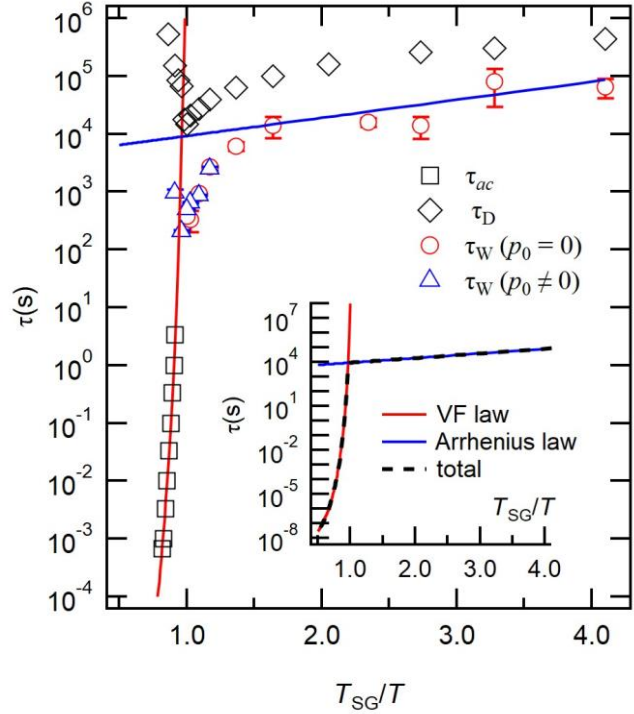

FIG. 10 Relaxation time $\tau$ as a function of $T_{SG}/T$. $\tau_{ac}$ is obtained by a least squares fitting ac-$\chi(\nu)$. $\tau_D$ and $\tau_W$ are obtained by numerical fittings of the decay curve of $M_{TRM}(t)$ with Debye and Weron relaxation functions, respectively. Inset shows $\tau_{VF}$, $\tau_{Arr}$, and $\tau_{tot}$ as a function of $T_{SG}/T$ represented by red and blue solid lines and a black dotted line, respectively.

Here we employ the Weron relaxation function (APPENDIX B) with a constant term $p_0$ to reproduce the isothermal decay curves of $M_{TRM}$. In the vicinity of the spin-glass transition at $T = T_{SG}$, parameters obtained by employing both the one-component ($p_0 = 0$) and two-component ($p_0 \neq 0$) models are compared in Figs. 9(a)–(c) and Fig. 10. The total moment $p_{tot} = p_0 + p_1$ increases rapidly in the vicinity of the spin-glass transition at $T_{SG} = 8.2$ K. The temperature variation of $p_{tot}$ is rather well described by $p_{tot}(T) = p_{tot}^0(1 - T/T_{SG})^\alpha$ at $T < T_{SG}$ with $p_{tot}^0 = 4.7(1)$ emu/mol and $\alpha = 0.77(5)$. Here, $p_{tot}^0$ is the TRM at $T = 0$. The exponent $\beta$ stays at ~0.5 below the transition, but it increases rapidly to ~1.0 at temperatures close to $T_{SG}$ (Fig. 9(b)). The coefficient $k$ decreases with increasing temperature and approaches zero at high temperatures (Fig. 9(c)). Note that in the limit of $k = 0$ and $\beta = 1.0$, the decay function asymptotically approaches a Debye function $p_1\exp(-t/\tau)$. The implication is that TRM relaxation above the spin-glass transition can be associated with a single spin flip with a weak magnetic correlation between spins. Moreover, the relaxation time $\tau$ increases with decreasing temperature and apparently follows an Arrhenius law with an activation energy $E_a \sim 100$ K at $T \sim T_{SG}$ and $E_a < 10$ K at $T \ll T_{SG}$. As shown in Fig. 10, at $T \sim T_{SG}$, the $\tau_W(T)$ and $\tau_D(T)$ curves approach the extrapolated curve of $\tau_{ac}(T)$ in accord with the Vogel–Fulcher law (Fig. 5). In contrast, above $T_{SG}$ the tendency of $\tau_{ac}(T)$ differs considerably. The discrepancy between $\tau_{ac}$ and $\tau_W$ is reconsidered below.

Here, we propose that the relaxation of the TRM may occur through either of two mechanisms: (1) the Vogel–Fulcher (VF) law and (2) an Arrhenius law with a small $E_a$, or a temperature-independent relaxation mechanism, such as a quantum mechanical process, i.e., tunneling through a potential barrier:

$$\frac{1}{\tau_{tot}} = \frac{1}{\tau_{VF}} + \frac{1}{\tau_{Arr}} \qquad (3)$$

(see Fig. 10). As shown above, it is plausible that $\tau_{VF}$ diverges exponentially at $T_0 = 7.14$ K. At $T \ll T_{SG}$ the temperature dependence of $\tau_{TRM}$ might be suppressed, and $\tau_{tot}$ may become nearly constant because of the presence of the second term in Eq. (3). It is generally established that the relaxation time $\tau$ scales with the number of relaxing units $N$ or effective volume $v$ of magnetic clusters responsible for magnetic relaxation according to a generalized expression such as $\tau = v^{1/\gamma}$ with $0 \leq \gamma \leq 1$ [28]. Consequently, the cluster size might be saturated at $T \ll T_{SG}$ in CoGa$_2$O$_4$, but it might not actually diverge anymore at $T = T_0$. In contrast, it is apparent in glass-forming liquids that the temperature variation of viscosity follows the VF law, i.e., viscosity diverges at $T_0$ below the glass transition temperature [40].

Figure 9(c) shows the temperature dependence of $k$, which is called interaction parameter. $k$ is related with the waiting time and a measure of collective nature of the interaction and associated with the non-extensivity parameter $q$ via the simple equation $q = (2k+1)/(k+1)$ [28] (see APPENDIX B). The value of $q$ varies in the range $1 < q < 2$ when $k > 0$. Theoretical findings reveal that the extensive–non-extensive critical phenomenon at $q = 5/3$ is expected to occur in complex physical systems such as glasses, polymers, and colloids [26]. In various metallic canonical spin-glass systems, measurement of the neutron spin echo has demonstrated that the $q(T)$ curves collapse in a curve with $q \sim 5/3$ at the spin-glass transition $T = T_{SG}$, whereas $q(T)$ asymptotically approaches $q = 2$ at $T = 0$ and reaches $q = 1$ at $T \sim 1.5 T_{SG}$ [25]. Surprisingly, the $q(T)$ obtained from the relaxation curves $M_{TRM}(t)$ of CoGa$_2$O$_4$ follow the $q(T)$ subtracted from the neutron spin echo spectra $S(Q, t)$ in canonical (metallic) spin-glasses, where $Q$ is the scattering vector and $t$ is in the range 0.01–10 ns.

### E. FT-IR spectra

A comparison of the FT-IR spectra of $ZnGa_2O_4$ and $CoGa_2O_4$ facilitated detection of the inversion effect on the infrared (IR) spectrum of $CoGa_2O_4$, which is sensitive to structure and the nature of bonding. Figure 11 shows the FT-IR spectra for $ZnGa_2O_4$ and $CoGa_2O_4$. The IR spectral peaks assigned theoretically to the $T_{1u}$ mode [41] were observed at the IR frequencies listed in Table II for $ZnGa_2O_4$. For the $ZnGa_2O_4$ synthesized in this study, the strong peaks at 582 and 439 cm$^{-1}$ corresponded well with previously calculated [41] and observed [42] values. Contrary to the case of $ZnGa_2O_4$, a shoulder and a broad peak were observed at 720 and 560 cm$^{-1}$, respectively, in the spectra of $CoGa_2O_4$ (Fig. 11). Compared with that of $ZnGa_2O_4$, the IR spectrum of $CoGa_2O_4$ shifted as expected toward higher frequencies when the mass difference between Co and Zn was considered. The shift was estimated to be detectable but small, e.g., $(m_{Zn}/m_{Co})^{1/2} = 1.05$. Actually, for the Zn and Co aluminates with $\eta < 0.055$, the shift of the IR peak position was negligibly small [43]. As shown in Fig. 11 and Table II, the IR peak at 620 cm$^{-1}$ of $CoGa_2O_4$ was shifted from the peak at 582 cm$^{-1}$ of $ZnGa_2O_4$ by a factor of 1.07, slightly larger than the 1.05 estimated above, whereas the IR peak at 434 cm$^{-1}$ might have been shifted toward lower frequency. In the largely inverted $CoGa_2O_4$, the additional structures of the IR spectra seemed to be related to the tetrahedrally coordinated $Ga^{3+}$ and the octahedrally coordinated $Co^{2+}$ ions. Whereas the mass difference between Co and Ga is relatively small, the phonon modes are rather strongly modified by the inversion, at least in the region 550–850 cm$^{-1}$. Note that in this study, the IR phonon modes below 400 cm$^{-1}$ of the spinel compounds were not detected because the lower frequency limit of the spectrometer was 398 cm$^{-1}$. These inversion-induced structures are also apparent in the inverse spinel $NiAl_2O_4$ with $\eta = 1$ (Table II) [44] and the partially inverted $ZnAl_2O_4$ nanoparticles with $\eta = 0.34$ [45]. The spectral intensity of the shoulder at 790 cm$^{-1}$ diminishes with decreasing $\eta$ in $ZnAl_2O_4$ nanoparticles annealed subsequently at various temperatures up to 1273 K [45, 46]. The shoulder has been claimed to be indicative of the presence of the inverted $Al^{3+}$ ion [45, 46]. It is plausible that in the IR spectrum of $CoGa_2O_4$, the shoulder at 810 cm$^{-1}$ is accompanied with the IR modes involving a $Ga^{3+}$ ion at the A-site.

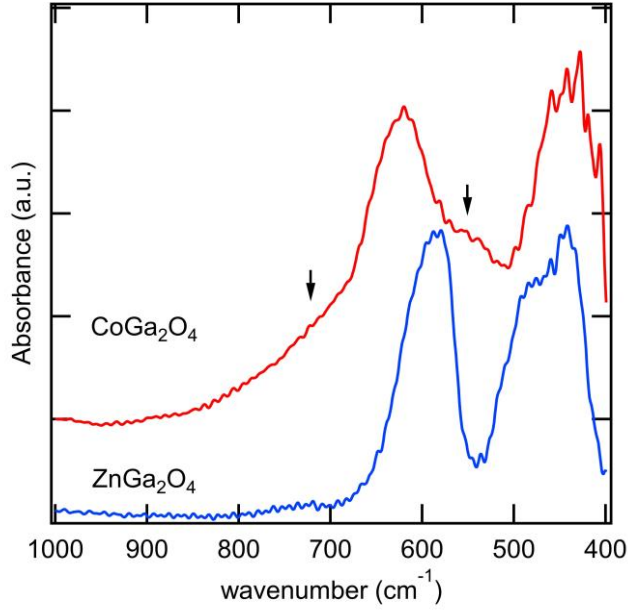

FIG. 11 FT-IR spectra for $CoGa_2O_4$ and the isostructural compound $ZnGa_2O_4$. Allows indicate the broad peak and shoulder observed at 560 and 720 cm$^{-1}$, respectively.

Table II Calculated and observed FT-IR mode frequencies (unit: cm$^{-1}$) for $ZnGa_2O_4$, $CoGa_2O_4$, and $NiAl_2O_4$.

| $ZnGa_2O_4$* | $ZnGa_2O_4$ | $ZnGa_2O_4$ | $CoGa_2O_4$ | $NiAl_2O_4$ |
|---|---|---|---|---|
| 580 | 570 | 582 | 720$^s$ | 810$^s$ |
| 429 | 420 | 439 | 620 | 715 |
| 342 | 328 |  | 560$^b$ | 600$^b$ |
| 175 | 175 |  | 434 | 505 |
| [41] | [42] | this work | this work | [44] |

*: Calculated values. *s*: shoulder. *b*: broad peak

### F. Specific heat

The magnetic state in $CoGa_2O_4$ was also characterized by specific heat measurements. The specific heat divided by temperature $C_{mol}/T$ exhibits a broad maximum at $T = 10$ K accompanied by a glass transition at $T_{SG} = 8.2$ K (Fig. 12(a)). To calculate the magnetic component of the specific heat $\Delta C_{mol}$ and magnetic entropy $S_{mag}$ we needed to estimate the lattice contribution of the specific heat $C_{lattice}$ for $CoGa_2O_4$. Theoretical considerations indicate that for $ZnGa_2O_4$ the partial phonon density of states (PDOS) for the heavy Zn and Ga ions is distributed mainly below 360 cm$^{-1}$, whereas for the light oxygen ion the partial PDOS is distributed above 360 cm$^{-1}$ [41]. At low frequency, the PDOS seems to exhibit a Debye-type parabolic frequency dependence that consists mainly of the contributions of Zn and Ga ions. We can apply the Einstein–Debye (ED) model to obtain $C_{lattice}$ for $MAl_2O_4$ (M = Zn, Co, Fe, Mn) [11] and for $CoAl_2O_4$ [16]. We followed the analytical procedures for aluminate oxides [11] and first made a fit of $C_{mol}/T$ for $ZnGa_2O_4$ to the ED function to obtain the Debye and Einstein temperatures (Table III). The ED function,

$$C = 3Ra_D x_D^{-3} \int_0^{x_D} \frac{x^4 e^x}{(e^x-1)^2} dT + R \sum_{k=1}^{2} a_{E_k} \frac{x_{E_k}^2 e^{x_{E_k}}}{\left(x_{E_k}-1\right)^2}, \quad (4)$$

consists of two terms and is based on the Debye–Einstein model with the three coefficients determined experimentally, i.e., the Debye temperature $\Theta_D$ and Einstein temperatures $\Theta_{E1}$ and $\Theta_{E2}$. The lattice degrees of freedom (LDF) for spinel compounds are distributed to one acoustic and two optical modes. Here, $x_i = \Theta_i/T$ ($i$ = D, $E_1$, $E_2$) are the reduced inverse temperatures, $a_i$ is the number of LDF, and $R$ is the gas constant. The Dulong–Petit rule predicts that in a three-dimensional lattice, the molar specific heat is $3rR$, which is reproduced asymptotically by the ED function at high temperatures ($T >> \Theta_D$, $\Theta_{E1}$, and $\Theta_{E2}$), where $r$ is the number of atoms per formula unit [47]. The factor $3r = 21$ is the number of LDF in a cubic spinel structure. The temperature variation of $C_{mol}/T$ for $ZnGa_2O_4$ was reproduced using the ED function with the coefficients shown in Table III. Compared with $\Theta_D$ = 286 K for $ZnAl_2O_4$ [11], the reduction of $\Theta_D$ for $ZnGa_2O_4$ by a factor 0.85 was in good agreement with the scaling factor of 0.82 derived analytically for a ternary compound [48]. In the ED function employed for the spinel compounds [11], the Einstein modes represented by $E_1$ and $E_2$ had 12 and 6 optical degrees of freedom, respectively, whereas the Debye mode had 3 acoustic degrees of freedom (Table III). It is quite reasonable to deduce for $ZnGa_2O_4$ that the $E_1$ and $E_2$ modes were due to oxygen and cation motions, respectively. In other words, the results for $\Theta_i$ and $a_i$ ($i$ = D, $E_1$, and $E_2$) shown in Table III are a coarse graining of the PDOS structure of $ZnGa_2O_4$ [41]. In the case of $ZnAl_2O_4$, the contributions of the motions of both the oxygen and aluminum atoms to the PDOS are primarily above 250 cm$^{-1}$, whereas the zinc motion contributes below 250 cm$^{-1}$. Above 700 cm$^{-1}$ and up to more than 1000 cm$^{-1}$, the PDOS is due to the motion of oxygen [49]. Therefore, the $E_1$ mode with $a_{E1}$ = 12 represents both oxygen and aluminum motions, whereas the $E_2$ mode with $a_{E2}$ = 6 is due to the motion of oxygen at high frequencies above 700 cm$^{-1}$. The simple ED function can reproduce the temperature variation of the specific heat, and the obtained coefficients are qualitatively consistent with the PDOS structures.

We needed a reasonable estimate of $C_{lattice}$ for CoGa$_2$O$_4$, because the phonon structure of CoGa$_2$O$_4$ was modified by the existence of the inverted cations (i.e., the octahedrally coordinated Co$^{2+}$ and the tetrahedrally coordinated Ga$^{3+}$), as suggested by the comparison of the IR spectra of the aluminates and gallates with various degrees of inversion. It is plausible, therefore, that the specific heat of the non-magnetic reference material ZnGa$_2$O$_4$ was not an appropriate $C_{lattice}$ for CoGa$_2$O$_4$. Assuming that the inversion effect on $\Theta_D$ could be neglected for CoGa$_2$O$_4$, we estimated the value of $\Theta_D$ for CoGa$_2$O$_4$ using the scaling factor $\Theta_D(CoGa_2O_4)/\Theta_D(ZnGa_2O_4) = 1.013$ [48]. We obtained a fairly good fit to the ED function above 90 K (Fig. 12(a)) and obtained the Einstein temperatures (Table III). A small failure in the curve fitting for $C_{mol}/T$ was apparent at $T > 200$ K. It is likely that the A-site substitution from Zn to Co resulted in a considerable shift of the phonon structure related to the motion of oxygen. The magnetic specific heat component $\Delta C_{mol}/T = [C_{mol}(CoGa_2O_4) - C_{lattice}(\Theta_D, \Theta_{E1}, \Theta_{E2})]/T$ and the magnetic entropy

$$S_{mag} = \int_{T_{min}}^{T} \frac{\Delta C_{mol}}{T} dT \quad (5)$$

were calculated as shown in Fig. 12(b). Here, $T_{min} \sim 1.9$ K was the lowest experimental temperature. $\Delta C_{mol}(T)$ followed the power law $\Delta C_{mol} = \delta T^{1.82}$ with respect to temperature, with $\delta = 54(1)$ mJK$^{-2.82}$ below $T = 8$ K (Fig. 12(b)), and it exhibited a maximum at $T = 11.5$ K. The exponent 1.82(1) was comparable to and somewhat larger than the typical value of 1.5 observed in spin glasses [50], but it was smaller than the reported values of 2.0 [11, 13], 2.1 [18], 2.23 [15], 2.33 [8], and 2.5 [12] for lightly inverted CoAl$_2$O$_4$ samples. With decreasing temperature, there was a remarkable upturn approximately proportional to $T^{-2}$, and a kink was apparent in the d$\Delta C_{mol}(T)$/d$T$ curve (Fig. 12(c)). Interestingly, the kink position coincided with $T_{SG}$. With a further decrease of temperature, the curve passed over a convex-upward plateau and then decreased monotonically. It is worth noting that $S_{mag}(T)$ seemed to reach a saturation value $S_{mag}$ of 7.5(9) JK$^{-1}$mol$^{-1}$ at temperatures as high as $T = 100$ K ($\gg T_{SG}$), and $S_{mag}(T)$ was suppressed compared with the experimentally obtained values of 13 at $T = 150$ K [11], 9.47 at $T = 100$ K [15], 10.2 at 60 K [18], 9.9 at 40 K [8], and 8.4 JK$^{-1}$mol$^{-1}$ at 100 K [16] for CoAl$_2$O$_4$ as well as for the theoretically expected value of $S_{spin} = R\ln(2S+1) = 11.52$ JK$^{-1}$mol$^{-1}$ based on the spin degrees of freedom for $S = 3/2$ of Co$^{2+}$.

As mentioned above, the magnetic measurement revealed that the effective magnetic moment for the Co$^{2+}$ at the B-site was considerably greater than the spin value of $p_{eff} = 3.87$ μ$_B$ for $S = 3/2$. This difference implies that an orbital moment was not completely quenched, and an intermediate ligand field scheme [51] was therefore appropriate for an octahedrally coordinated Co$^{2+}$, that is, the crystal field and spin-orbit coupling energies seemed to be comparable. The Co$^{2+}$ (3$d^{\,7}$) state was described using a fictitious angular moment $\zeta = 1$ that coupled to the spin moment $S = 3/2$ [51]. Spin-orbit coupling resulted in the formation of a (2$\zeta$+1)(2$S$+1) manifold of $J' = \zeta + S$ consisting of three states of $J' = 5/2, 3/2$, and $1/2$. The ground state was the doublet of $J' = 1/2$. The energy separation between the ground and low-lying $J' = 3/2$ states was estimated to be approximately 580 K for the octahedral Co$^{2+}$ ion. This energy difference was a result of spin-orbit coupling $\lambda'(\zeta \cdot S)$, where $\lambda'$ is an effective spin-orbit parameter. Therefore, the estimated specific heat involved with the excited states seemed to be negligibly small in the temperature range $T < 100$ K. For CoGa$_2$O$_4$, the magnetic entropies of the tetrahedral and the octahedral Co$^{2+}$ ions were $(1 - \eta)R\ln(4)$ and $\eta R\ln(2)$, respectively. The experimental value of $S_{mag} = 7.5$ JK$^{-1}$mol$^{-1}$ at $T = 100$ K was, within experimental error, equal to $(1 - \eta)R\ln(4) + \eta R\ln(2) = 7.70$ JK$^{-1}$mol$^{-1}$. For the octahedrally coordinated Co$^{2+}$ ion, the doublet grand state stabilized by the effective spin-orbit coupling was also apparent in the garnet NaCa$_2$Co$_2$V$_3$O$_{12}$, which exhibited an antiferromagnetic transition at $T_N = 6.4$ K [52], as well as the silver delafossite Ag$_3$Co$_2$SbO$_6$ with $T_N = 21.2$ K [53].

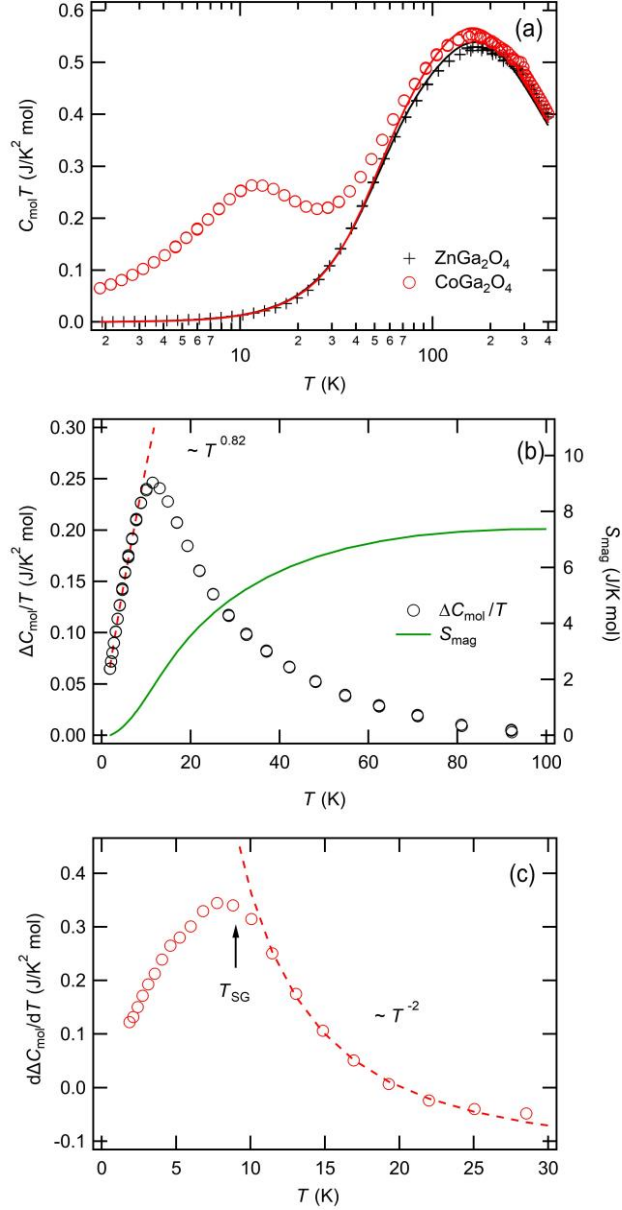

FIG. 12 (a) Temperature variations of $C_{mol}/T$ for $CoGa_2O_4$ and the isostructural compound $ZnGa_2O_4$. Red and black lines represent $C_{lattice}/T$ obtained by least squares fitting to the Einstein–Debye model for $CoGa_2O_4$ and $ZnGa_2O_4$, respectively. (b) Black open circles show the magnetic component $\Delta C_{mol}/T = [C_{mol}(CoGa_2O_4) - C_{lattice}]/T$. The magnetic entropy $S_{mag}$ is also depicted. The dashed line represents a power law with respect to temperature $\Delta C_{mol}/T = 0.054 T^{0.82}$. (c) The temperature derivative of $\Delta C_{mol}$ as a function of temperature. Arrow indicates the spin-glass transition point $T_{SG}$. The dashed red line shows $T^2$-variation above $T_{SG}$.

Table III  Lattice degrees of freedom $a_i$ ($i = D$, $E_1$, and $E_2$) employed in Ref. 11 and Debye and Einstein temperatures for $CoGa_2O_4$, $ZnGa_2O_4$, and related spinel compounds.

| Mode* | D(A) | $E_1$(O) | $E_2$(O) | |
|---|---|---|---|---|
| $a_i$ | 3 | 12 | 6 | 21 in total |
| Temperature | $\Theta_D$ (K) | $\Theta_{E1}$ (K) | $\Theta_{E2}$ (K) | Ref. |
| $CoGa_2O_4$ | 247** | 634 | 301 | this work |
| $ZnGa_2O_4$ | 244 | 672 | 310 | this work |
| $CoAl_2O_4$ | 301 | 530 | 886 | [11] |
| $ZnAl_2O_4$ | 286 | 543 | 1244 | [11] |

* A and O in parentheses represent acoustic and optical modes, respectively.

** Estimated by a scaling factor of $\Theta_D(CoGa_2O_4)/\Theta_D(ZnGa_2O_4) = 1.013$ [48].

## IV. DISCUSSION AND CONCLUSIONS

We demonstrated by measurements of the dc- and ac-magnetic susceptibilities and TRM that cluster glass is established below the transition at $T_{SG} = 8.2$ K in $CoGa_2O_4$. In addition, the realization that TRM was a macroscopic phenomenon proved that there is a mechanism of magnetic relaxation in the glass state. In the following discussion, we summarize the results for $CoGa_2O_4$ that were different from or comparable to previously reported characteristics of other spin and cluster glasses.

The cationic configuration in $CoGa_2O_4$ is close to that of a random spinel characterized by a degree of inversion η = 2/3. The Rietveld refinement revealed that the inversion of our polycrystal sample was 0.664(8). The implication is that both the A-site and B-site occupancies of the magnetic $Co^{2+}$ ion, $g_A$(Co) and $g_B$(Co), respectively, were greater than the percolation threshold $x_c^{AB} = 0.227(3)$, but they were smaller than the A-site and B-site thresholds $x_c^A = 0.429(3)$ and $x_c^B = 0.390(3)$ [54], respectively. The $Co^{2+}$ ions with $S = 3/2$ were distributed in both the sites and form macroscopic magnetic clusters if the nearest neighbor super-exchange interactions $J_{AB}$ and $J_{BB}$ are anticipated between spins. These circumstances eliminate the bond frustration expected to exist in the A-site spinel antiferromagnet and also induce the cluster glass state. The octahedrally coordinated $Co^{2+}$ ion is known to display weak Jahn–Teller activity because of the orbital (configurational) degree of freedom. Because in $CoGa_2O_4$ the $g_B(Co)$ is less than $x_c^B$, it seems theoretically possible that a cooperative Jahn–Teller transition is not realized. In actuality, the crystal structure of $CoGa_2O_4$ is always a cubic spinel.

The relaxation rate $1/\tau_D$ and the temperature derivative of the $M_{TRM}$ are sensitive to the spin-glass transition at $T_{SG}$. They exhibit sharp peaks at $T_{SG}$ (Figs. 6 and 7(b)). It is worth to note that the frequency dependent ac-susceptibility is observed below $T \sim 50$ K (Fig. 4), which seems to suggest that $CoGa_2O_4$ undergoes a short-range magnetic order. Since the imaginary component $\chi''(T)$ does not show an anomaly at $T \sim 50$ K, it is plausible that an antiferromagnetic cluster embedded in paramagnetic spins is developed at $T \gg T_{SG}$ [55]. This speculation is supported by the discrepancy between the temperature variations of $\tau_{ac}$ and $\tau_D$ observed at $T > T_{SG}$, as shown in Fig. 10, that is, a two-component model can be employed for describing the magnetic state of $CoGa_2O_4$. Therefore, more plausibly, an inhomogeneous magnetic phase, e.g., a Griffiths phase emerges below $T = |\theta|$ where $\theta$ is the Weiss temperature obtained in the paramagnetic state, as speculated to be realized in $CoAl_{2-x}Co_xO_4$ at $x > 0.9$ [56]. These situations for $CoGa_2O_4$ remind us that in a $Co_3O_4$ nanoparticle system a spin glass-like transition occurs at $T = 10$ K, while an antiferromagnetic transition is indicated at $T = 32$ K close to the Néel point for a bulk sample [57]. Below the Néel point a ferromagnetic component arises and develops with decreasing temperature. Interparticle magnetic interactions realized in the antiferromagnetic nanocrystals randomly oriented brings about the spin glass-like transition.

In $CoGa_2O_4$ the saturation tendency of the relaxation time (i.e., cluster volume) observed at low temperatures $T < T_{SG}$ reflects the fact that in the A-site frustrated antiferromagnet $CoAl_2O_4$, the magnetic correlation length remains on the order of nanometers at well below the Néel temperature [17, 18]. The absence or suppression of long-range ordering in $CoGa_2O_4$ might be due to the large degree of inversion. Concomitantly, the $Co^{2+}$-ion occupying the octahedral B-site might have brought about the formation of the magnetic cluster. In fact, $M_{TRM}$ is enhanced as $\eta$ increases with increasing $x$ in $CoAl_{2-x}Ga_xO_4$ [39]. On the other hand, the suppression of long-range magnetic ordering in $CoAl_2O_4$ seems to result from the proximities to the NSS boundary of $J_2/J_1 = 1/8$ and the NSG boundary of $\eta_c = 0.08$ (Fig. 1(b)).

While relaxation time τ(T) reflects the inherent characteristics in $CoGa_2O_4$, apparent magnetic behaviors might be quite universal as those expected to be observed in a spin or cluster glass as follows. At $T = T_{SG}$ the non-extensive parameter is $q \sim 5/3$ for $CoGa_2O_4$, where the extensive–nonextensive transition is expected to occur [26], we observed (i) the SG transition accompanied by the critical slowing down of spin fluctuations, (ii) the form of relaxation changing from a Debye to a non-exponential form, and (iii) thermoremanent magnetization $M_{TRM}(0)$ emerging rapidly. The presented comprehensive magnetic investigation for the random spinel magnet $CoGa_2O4$ might also provide insight for understanding magnetic properties for the frustrated antiferromagnet $CoAl_2O_4$ assigned to be in the vicinity of the NSS and NSG boundaries and facilitates further investigations.

## APPENDIX A (Spin-glass transition at $T_{SG}$ and critical exponent $zv'$)

As can be seen in Fig. A1, the relaxation time $\tau_{ac}$ for $CoGa_2O_4$ shows a conventional power-law divergence of the critical slowing down at $T = T_{SG}$,

$$\tau_{ac}(T) = \tau_0 \left( \frac{T_f - T_{SG}}{T_{SG}} \right)^{-zv'}, \qquad (A1)$$

where $\tau_0$ is a characteristic relaxation time. Here we obtain $\tau_0 = 2.5(6) \times 10^{-10}$ s, $zv' = 9.8(1)$, and $T_{SG} = 8.2$ K by the least squares method. The value of $\tau_0$ corresponds well with that obtained above by the fitting to the VF law $2.86 \times 10^{-10}$ s. The power exponent of $zv' = 9.8$ is comparable to the values in the range 4–12 found for canonical spin-glass systems [16]. Generally, the correlation length $\xi(T)$ diverges as $\sim (T - T_c)^{-v'}$, and the relaxation time as $\tau \sim \xi^z \sim (T - T_c)^{-zv'}$ when the temperature approaches the transition point, $T_c$.

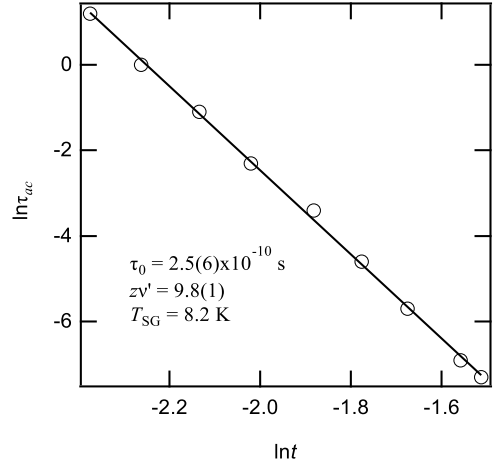

FIG. A1 Logarithmic plot of relaxation time $\tau_{ac}$ as a function of reduced temperature $t = (T_f − T_{SG})/T_{SG}$. Solid line represents the results of a least squares fit to the data.

**APPENDIX B (nonexponential relaxation functions)**

The cluster of concern here is a microscopic and mesoscopic unit consisting of a large number of relaxing entities which can be characterized by its volume or relaxation time. The cluster volume is closely related to a coherence length evaluated directly by microscopic probes, while relaxation time in a cluster is significantly different from its surroundings. Relaxation in glasses, which are typical complex physical systems, is a stochastic process that depends both on the geometric characteristics of the cluster, such as its morphology or fractal dimension, and on the collective nature of the interaction [28]. Generally, liquid and glass states are structurally quite similar. The most important and representative quantity between liquid and glassy states seems to be the degree of the relaxation time. In contrast with the logarithmic function $m_0 + S \cdot \ln(t)$ used generally for spin glass systems, the following relaxation functions explicitly contain the relaxation time $\tau$.

**1) Kohlrausch (stretched exponential) function**

The long-tail relaxation is well reproduced by the Kohlrausch (stretched exponential) form

$$\phi(t) = \exp\left[-\left(\frac{t}{\tau}\right)^{\beta}\right] \qquad (B1)$$

for $0 < \beta < 1$. This function is asymptotically a logarithmic and Debye function at $\beta \rightarrow 0$ and 1, respectively.

**2) Ogielski function**

Monte-Carlo calculations for a 3D- $\pm J$ Ising spin-glass model were conducted by Ogielski [58], who obtained the spin autocorrelation function of the spin-glass state. This function is phenomenologically expressed by

$$\phi(t) = t^{-x}\exp\left[-\left(\frac{t}{\tau}\right)^{\beta}\right] \tag{B2}$$

where the exponent $x$ is in the range 0–0.5, and the relaxation time $\tau$ diverges at $T_{SG}$.

**3) Weron's generalized probabilistic relaxation function**

To describe dielectric relaxation consistent with experimentally established findings and based on purely stochastic theories, Weron proposed a universal relaxation function in a power-law form [24]:

$$\phi(t) = \left[1 + k\left(\frac{t}{\tau}\right)^{\beta}\right]^{-\frac{1}{k}} \tag{B3}$$

for $k > 0$ and $0 < \beta < 1$, where $k$ and $\beta$ are mathematical macroscopic parameters. The parameter $k$ is called the interaction parameter and is related to the waiting time $t_w$, and "β represents macroscopically the 'fractal' geometry and dynamic nature of the relaxation dynamics" [28]. In the case of $k \rightarrow 0$, the stretched exponential form, Eq. (B1), is recovered. Later it was revealed that the dipolar relaxation function $\phi_{\gamma,q}(t)$ was derived for a cluster model based on the generalized maximum Tsallis nonextensive entropy principle [28], whereas it is assumed quite generally that the relaxation time $\tau$ of a volume $v$ made of $N$ relaxing elements scales as $\tau = v^{1/\gamma}$ with $0 < \gamma < 1$. The Tsallis generalized nonextensive entropy [26, 27] is formulated as follows:

$$S_q = k_B \frac{1 - \int_0^{\infty} p^q(x)dx}{q-1}. \tag{B4}$$

Eq. B4 consistently encloses the Boltzmann-Gibbs entropy $S_{BG} = -k_B \int_0^{\infty} p(x)\ln p(x)dx$ as an asymptotic form when the non-extensive parameter $q \rightarrow 1$. Here $p(x)$ is a distribution function in the sense of a probabilistic point of view, and $x$ is, for example, a correlation length $\xi$ in a dielectrically or magnetically correlated system or cluster volume $v$ of an aggregated cluster system. Remarkably, asymptotic power laws of the response function $f_{\gamma,q}(t) = -\frac{d\phi_{\gamma,q}}{dt}$ obtained in both the limits $t \to 0$ and $t \to \infty$ are consistent with those derived from the Weron function Eq. (B3) [28]. This coincidence brings about a simple relation between the Tsallis parameter $q$ in Eq. (B4) and the stochastic parameter $k$ in Eq. (B3), i.e., $k = (q-1)/(2-q)$, in addition to $\beta = \gamma$. Moreover, it gives physical meanings to $k$ and $\beta$, which were introduced as purely stochastic parameters by Weron [24].

**APPENDIX C  Effects of aging on the TRM relaxation curve**

It is known in spin glasses [20, 36] that the relaxation rate with respect to logarithmic time, which is defined as

$$R(t) = -d\left(\frac{M_{TRM}}{H_{FC}}\right)/d\ln t, \quad (C1)$$

exhibits a maximum at $t \sim t_w$. The exhibition of this maximum is a kind of aging behavior. In $CoGa_2O_4$, however, $R(t)$ does not show a maximum, and there is some $t_w$ dependence of $R(t)$, but the dependence is quite small in the range of $50 \leq t_w \leq 600$ s (Fig. C1). In the TRM relaxation curves for $CoGa_2O_4$ taken after the waiting duration $t_w = 300$ s, the fitting procedure and extracted parameters seem not to be seriously affected by aging effects.

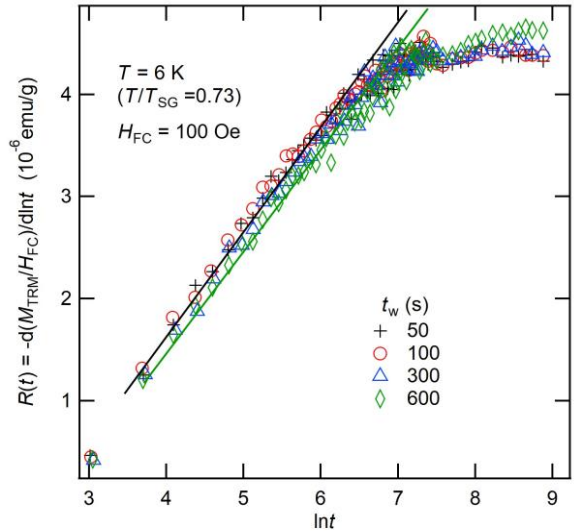

FIG. C1 Relaxation rate $R(t)$ as a function of logarithmic time $\ln t$ at various waiting times. Solid line is a guide to facilitate visualization.

## ACKNOWLEDGEMENTS

This work was partially supported by a Grant-in-Aid for Scientific Research, KAKENHI (25246013, 16K13999, 17H03404, and 20H01851) and the Research Institute for Science and Technology of Tokyo Denki University (Grant Number: Q19K-02). The synchrotron radiation experiments were performed on the BL15XU beam line at SPring-8 with the approval of the Japan Synchrotron Radiation Research Institute (2015B4505, 2016B4505, and 2018A4507). The authors are grateful to Hiroaki Mamiya at the National Institute for Materials Science for valuable comments and suggestions on the ac-susceptibility measurement.